\documentclass[twocolumn,showpacs,aps,prl,superscriptaddress]{revtex4}

\usepackage[T1]{fontenc}
\usepackage{graphicx}
\usepackage{dcolumn}
\usepackage{amsmath}
\usepackage{epsfig}
\usepackage{lipsum}

\input babarsym

\newcommand{\BABARPubYear}    {14}
\newcommand{\BABARPubNumber}  {004}

\newcommand{\SLACPubNumber} {16118}

\def\figurebox#1#2#3{%
    \def\arg{#3}%
    \ifx\arg\empty
    {\hfill\vbox{\hsize#2\hrule\hbox to #2{\vrule\hfill\vbox to #1{\hsize#2\vfill}\vrule}\hrule}\hfill}%
    \else
    {\hfill\epsfbox{#3}\hfill}%
    \fi}

\begin{document}

\preprint{\babar-PUB-\BABARPubYear/\BABARPubNumber} 
\preprint{SLAC-PUB-\SLACPubNumber}

\begin{flushleft}
\babar-PUB-\BABARPubYear/\BABARPubNumber\\
SLAC-PUB-\SLACPubNumber
%arXiv:\LANLNumber\ [hep-ex]\\[10mm]
\end{flushleft}

\def\Lambdabar {\kern 0.2em\overline{\kern -0.2em \Lambda}{}\xspace}

\def\Lcp {\Lambda^+_c\xspace}
\def\Lcm {\Lambdabar^-_c\xspace}
\def\Lc  {\Lambda_c\xspace}
\def \BLcpKK {\ensuremath{\Bzb \to \Lcp \antiproton \Km \Kp}\xspace}
\def \BLcpKzKz {\ensuremath{\Bzb \to \Lcp \antiproton \Kzb \Kz}\xspace}
\def \BLcpphi {\ensuremath{\Bzb \to \Lcp \antiproton \phi}\xspace}
\def \BLcppipi {\ensuremath{\Bzb \to \Lcp \antiproton \pim \pip}\xspace}
\def \BLcpKpi {\ensuremath{\Bzb \to \Lcp \antiproton \Km \pip}\xspace}
\def\mb        {\mbox{$m_{\B}$}\xspace}

\title{
{\large \bf
\boldmath
Observation of the baryonic decay $\BLcpKK$
\unboldmath
} 
}

% Dummy author list; contact PubBoard Chair for current author list
%% author list as of 05-May-2014 (311 authors)
%
\author{J.~P.~Lees}
\author{V.~Poireau}
\author{V.~Tisserand}
\affiliation{Laboratoire d'Annecy-le-Vieux de Physique des Particules (LAPP), Universit\'e de Savoie, CNRS/IN2P3,  F-74941 Annecy-Le-Vieux, France}
\author{E.~Grauges}
\affiliation{Universitat de Barcelona, Facultat de Fisica, Departament ECM, E-08028 Barcelona, Spain }
\author{A.~Palano$^{ab}$ }
\affiliation{INFN Sezione di Bari$^{a}$; Dipartimento di Fisica, Universit\`a di Bari$^{b}$, I-70126 Bari, Italy }
\author{G.~Eigen}
\author{B.~Stugu}
\affiliation{University of Bergen, Institute of Physics, N-5007 Bergen, Norway }
\author{D.~N.~Brown}
\author{L.~T.~Kerth}
\author{Yu.~G.~Kolomensky}
\author{M.~J.~Lee}
\author{G.~Lynch}
\affiliation{Lawrence Berkeley National Laboratory and University of California, Berkeley, California 94720, USA }
% ...
\collaboration{The \babar\ Collaboration}
\noaffiliation

%\date{\today}% It is always \today, today, but you may specify any date with \date.

\begin{abstract}
We report the observation of the baryonic decay \BLcpKK using a data sample of $471 \times 10^6$ \BB pairs produced in \epem annihilations at $\sqrt{s}=10.58 \gev$. This data sample was recorded with the $\babar$ detector at the \pep2 storage ring at SLAC. We find $\BR\left(\BLcpKK \right) = \left(2.5 \pm 0.4_{(\text{stat})} \pm 0.2_{(\text{syst})}\pm 0.6_{\BR\left(\Lcp\right)} \right) \times 10^{-5}$, where the uncertainties are statistical, systematic, and due to the uncertainty of the $\Lcp \to \proton \Km \pip$ branching fraction, respectively. The result has a significance corresponding to 5.0 standard deviations, including all uncertainties.
For the resonant decay \BLcpphi, we determine the upper limit $\BR\left( \BLcpphi \right) <1.2 \times 10^{-5}$ at $90\%$ confidence level.
\end{abstract}


\maketitle

About $7\%$  of all \B mesons decay into final states with baryons \cite{ref:PDG}. Measurements of the branching fractions for baryonic \B decays and studies of the decay dynamics, e.g., the fraction of resonant subchannels or the possible enhancement in the production rate at the baryon-antibaryon threshold seen in some reactions \cite{delAmoSanchez:2011gi,Aubert:2010zv}, can provide detailed information that can be used to test phenomenological models \cite{Carlstone:1969tr,Rosner:2003bm,Suzuki:2006nn}. Studying baryonic \B decays can also allow a better understanding of the mechanism of these decays and, more generally, of the baryon production process.

In this paper we present a measurement of the branching fraction for the decay \BLcpKK. Throughout this paper, all decay modes include the charge conjugate process. No experimental results are currently available for this decay mode. However, the related decay \BLcppipi has been observed with a branching fraction $\BR(\BLcppipi) = (1.17\pm0.23) \times 10^{-3}$ \cite{ref:PDG}. The main difference between the decay presented here and \BLcppipi is that there are fewer kinematically accessible resonant subchannels for \BLcpKK. The heavier mass of the $\s$ quark suggests a suppression factor of about $1/3$ \cite{Andersson:1983ia}, which is consistent with the observed suppression of $\Bzb \to \Dz \Lambda \Lambdabar$ relative to $\Bzb \to \Dz \proton \antiproton$ \cite{cvoss2372}. However, the \BLcpKK and \BLcppipi decay processes are described by different Feynman diagrams, and this simple expectation might not hold.

The analysis is based on an integrated luminosity of $429 \invfb$ \cite{Lees:2013rw} of data collected at a center-of-mass energy equivalent to the $\FourS$ mass, $\sqrt{s}=10.58 \gev$, with the $\babar$ detector at the \pep2 asymmetric-energy \epem collider at SLAC, corresponding to $471 \times 10^6$ \BB pairs. Trajectories of charged particles are measured with a five-layer double-sided silicon vertex tracker and a 40-layer drift chamber, operating in the $1.5$\,T magnetic field of a superconducting solenoid. Ionization energy loss measurements in the tracking chambers and information from an internally reflecting ring-imaging detector provide charged-particle identification \cite{ref:PID}. The $\babar$ detector is described in detail elsewhere \cite{ref:NIM,ref:NIM2}. Monte Carlo (MC) simulations of events are used to study background processes and to determine signal efficiencies. The simulations are based on the {\texttt EvtGen} \cite{ref:EvtGen} event generator, with the {\texttt Geant4} \cite{ref:geant} suite of programs used to describe the detector and its response. The \BLcpKK and $\Lcp \to \proton \Km \pip$ final states are generated according to four-body and three-body phase space, respectively.\\

We reconstruct $\Lcp$ baryons in the decay mode $\Lcp \to \proton \Km \pip$. For the \B meson reconstruction, we combine the $\Lcp$ candidate with identified $\antiproton$, $\Km$, and $\Kp$ candidates and fit the decay tree to a common vertex constraining the $\Lcp$ candidate to its nominal mass. We require the $\chi^2$ probability of the fit to exceed $0.001$. To suppress combinatorial background, we require the $\Lcp$ candidate mass to lie within approximately two standard deviations ($\pm 10\mevcc$) in the expected resolution from the nominal $\Lcp$ mass.

\begin{figure}[t]
  \begin{center}
    \includegraphics[width=0.49\textwidth]{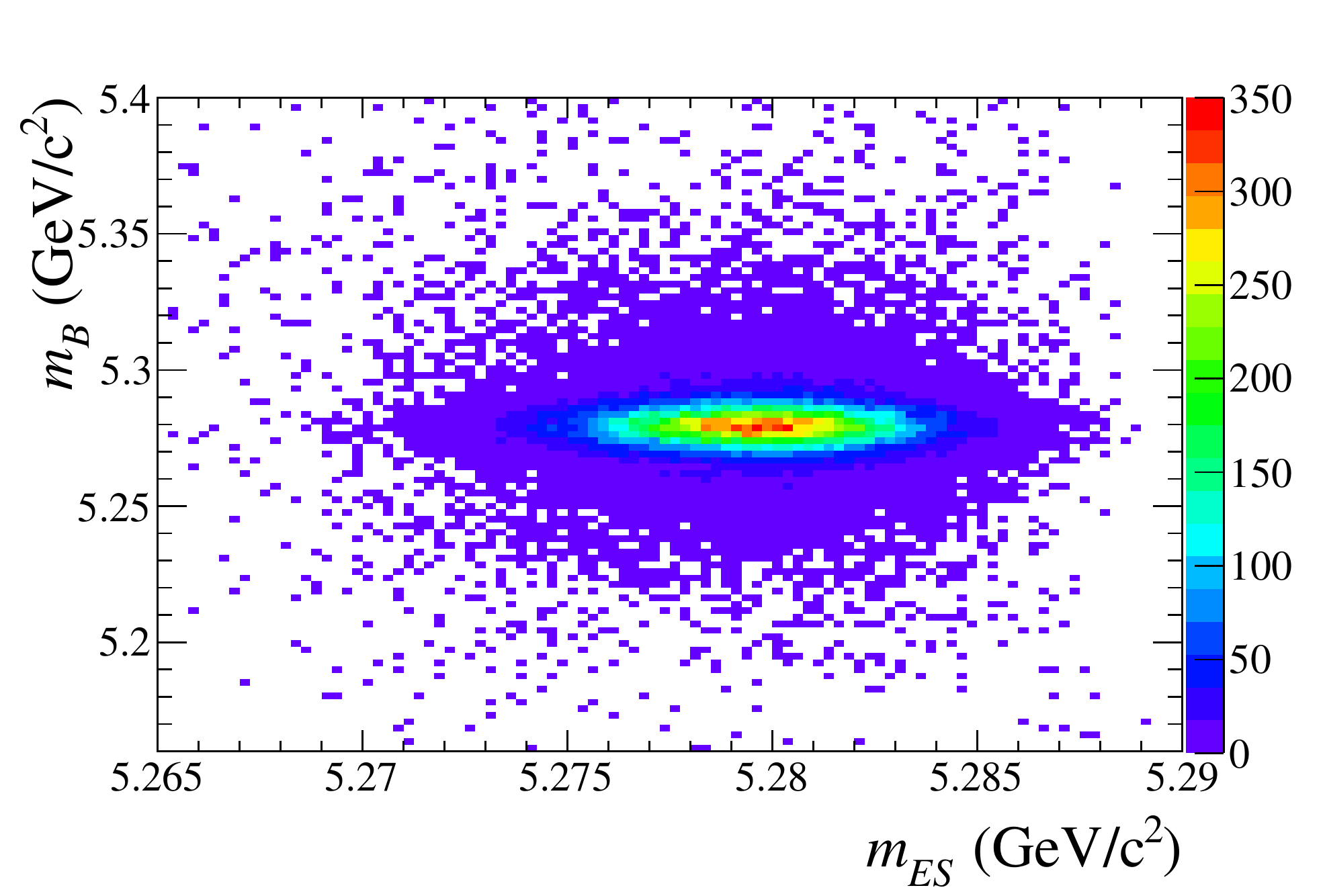}
  \end{center}
  \caption{The $\mb$ vs $\mes$ distribution for correctly reconstructed simulated signal events.}
  \label{fig:scatter}
\end{figure}

We determine the number of signal candidates with a two-dimensional unbinned extended maximum likelihood fit to the \B meson candidate invariant mass, $\mb$, and the energy-substituted mass, $\mes$, defined as 
\begin{equation}
  \mes = \sqrt{\left( \frac{s/2 +\vec{p}_{\B}  \cdot \vec{p}_0}{E_0} \right)^2 - \vec{p}_{\B}^{\,2}},
\end{equation}

\noindent where the \B momentum vector, $\vec{p}_{\B}$, and the four-momentum vector of the $\epem$ system, $\left(E_0,\vec{p}_0\right)$, are measured in the laboratory frame. For correctly reconstructed \B decays, $\mb$ and $\mes$ are centered at the nominal \B mass. The correlation between $\mb$ and $\mes$ in simulated signal (Fig.~\ref{fig:scatter}) and background events is approximately zero and not significant. It can be neglected in this analysis. For signal events, the shape of the \mes distributions is described by the sum $f_{2G}$ of two Gaussian functions, as is the \mb distribution. The means, widths, and relative weights in the four Gaussians are determined using simulated events and are fixed in the final fit. Background from other \B meson decays and continuum events $(\epem \to \qqbar,\, \q=\u,\,\d,\,\s,\,\c)$ is modeled using an ARGUS function \cite{ref:argus}, $f_\text{ARGUS}$, for $\mes$ and a first-order polynomial, $f_\text{poly}$, for $\mb$.

The fit function is defined as
\begin{equation}
\begin{split}
 f_{\text{fit}} &= N_{\rm sig} \cdot \mathcal{S(\mes, \mb)} + N_{\rm bkg} \cdot \mathcal{B(\mes, \mb)}\\
&= N_{\rm sig} \cdot f_\text{2G}(\mes)\cdot f_\text{2G}(\mb) \\
&\quad + N_{\rm bkg} \cdot f_\text{ARGUS}(\mes)\cdot f_\text{poly}(\mb),
\end{split}
\end{equation}

\noindent where $N_{\rm sig}$ and $N_{\rm bkg}$ are the number of signal and background events, respectively, with $\mathcal{S}$ and $\mathcal{B}$ the corresponding probability density functions (PDFs). The extended likelihood function is
\begin{equation}
\begin{split}
L(N_{\rm sig},N_{\rm bkg}) = \qquad \qquad \quad\\
 \frac{e^{-(N_{\rm sig}+N_{\rm bkg})}}{N!} \prod_{i = 1}^{N} [&N_{\rm sig} \mathcal{S}_i(\mes_{i}, m_{\B i}) \\&+ N_{\rm bkg} \mathcal{B}_i(\mes_i, m_{\B i})],
\end{split}
\end{equation}

\noindent where $i$ denotes the $i$th candidate and $N$ is the total number of events in the fit region. The fit region is defined by the intervals $5.2 \gevcc < \mb < 5.55 \gevcc$ and $5.2 \gevcc < \mes < 5.3 \gevcc$.

\begin{figure}[t]
  \begin{center}
	\includegraphics[width=0.49\textwidth]{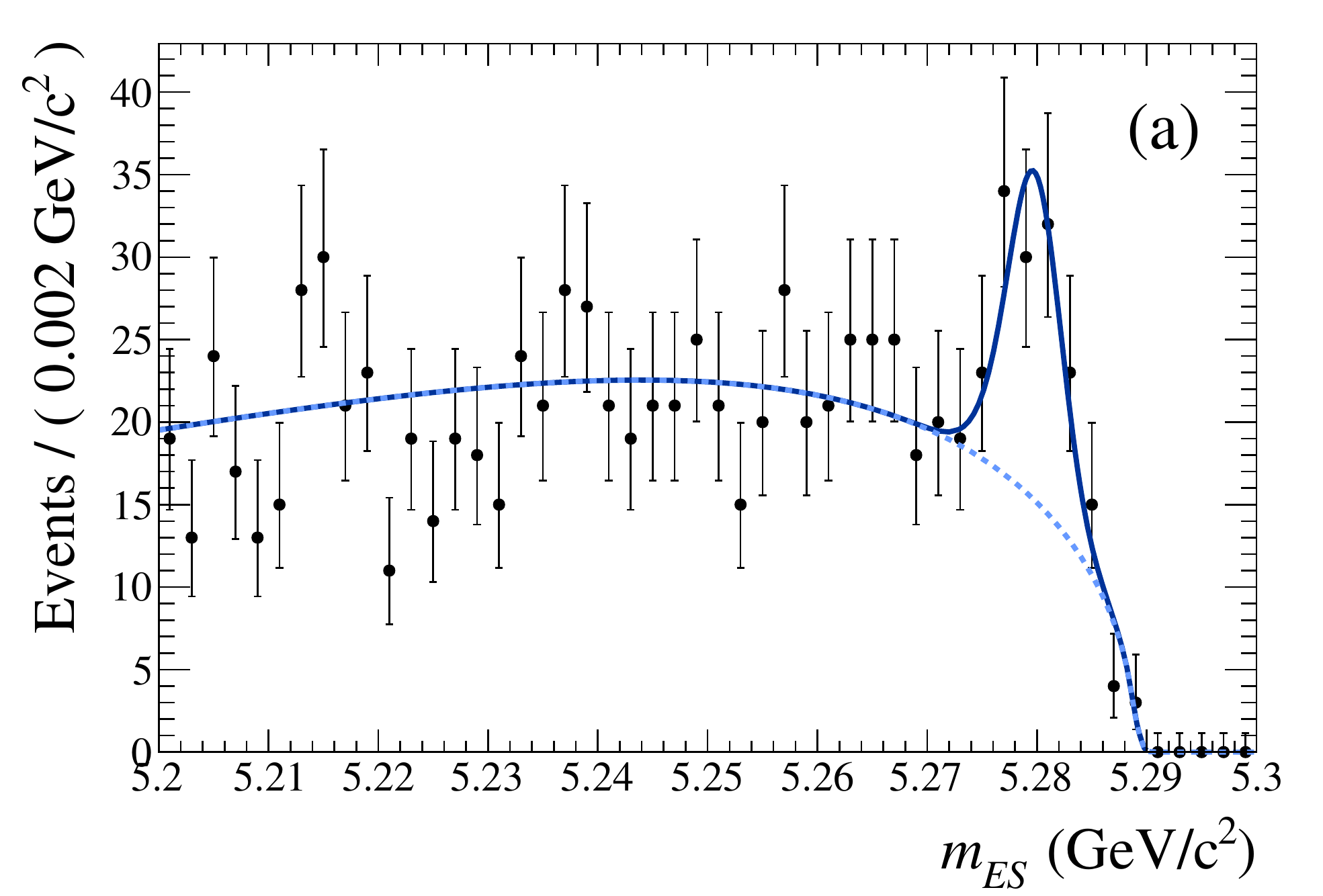}
	\hfill
	\includegraphics[width=0.49\textwidth]{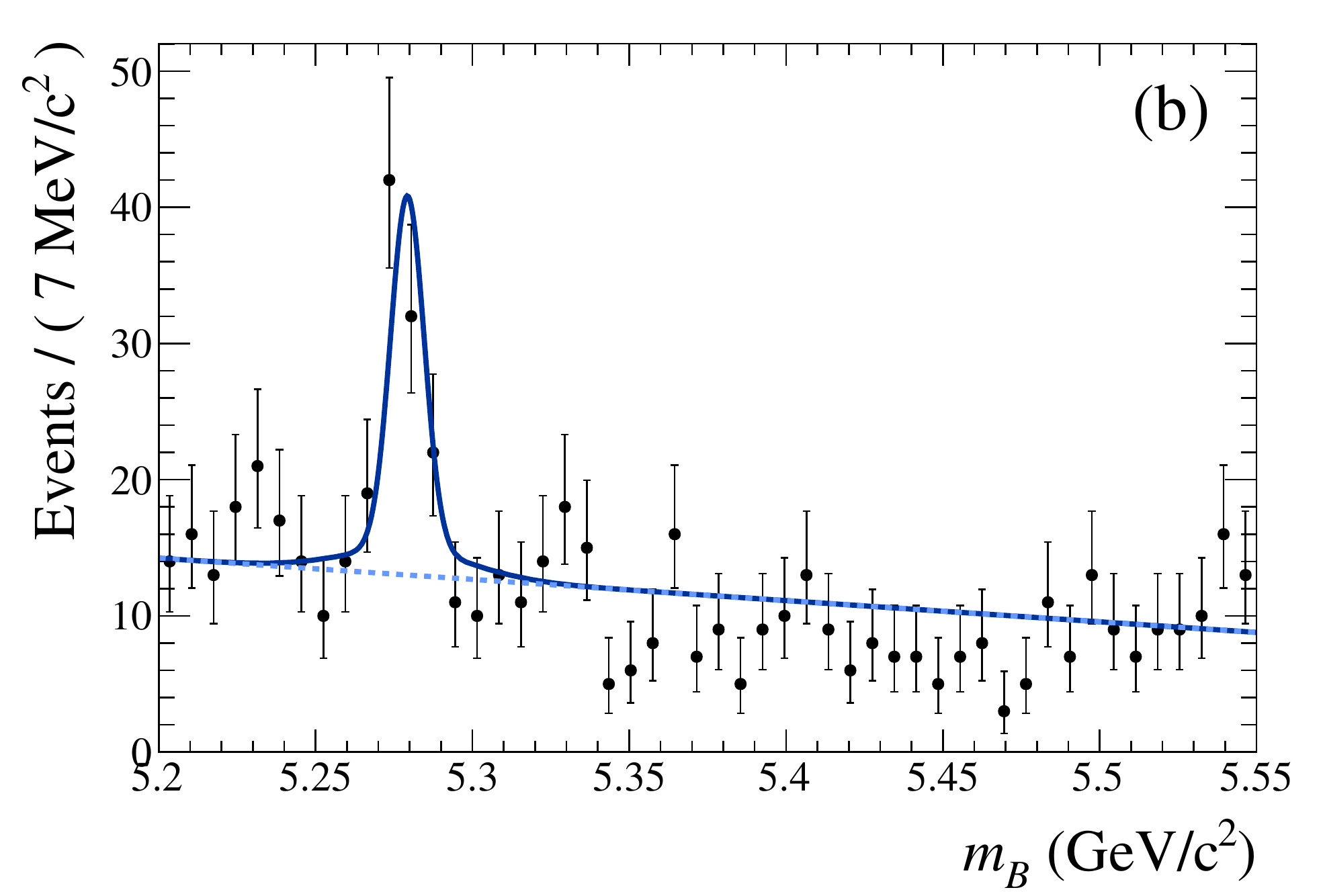} 
  \end{center}
  \caption{Data (points with statistical uncertainties) and projections of the maximum likelihood fit (solid curves) for \BLcpKK candidates. The dashed curves show the projections of the PDF for background events. (a) Results for $\mes$, with the requirement $5.26 \gevcc \leq \mb \leq 5.30 \gevcc$. (b) Results for $\mb$, with the requirement $5.275 \gevcc \leq \mes \leq 5.285 \gevcc$.}
  \label{fig:data}
\end{figure}

Figure \ref{fig:data} shows the one-dimensional projections of the fit results onto the $\mes$ and $\mb$ axes in comparison with the data. Clear signal peaks at the \B meson mass are visible. We find $N_\text{sig} = 66 \pm 12$, where the uncertainty is statistical only. The statistical significance $S$ of the signal is determined from the ratio of the likelihood values for the best-fit signal hypothesis, $L_\text{sig}$, and the best fit with no signal included, $L_0$, $S = \sqrt{-2 \ln(L_0 / L_\text{sig})}$, corresponding to $5.4$ standard deviations.

The efficiency to reconstruct signal events depends on the baryon-antibaryon invariant mass. Therefore, to determine the $\BLcpKK$ branching fraction, we divide the data into two regions. Region I is defined as $3.225 \gevcc < m_{\Lcp \antiproton} \leq 3.475 \gevcc$ and region II as $3.475 \gevcc < m_{\Lcp \antiproton} \leq 4.225 \gevcc$. The results are shown in Table \ref{tab:eff}. To determine an upper limit on the branching fraction for the decay \BLcpphi, we do not divide the events into regions of $m_{\Lcp \antiproton}$. Instead we use only events in the $\phi$ signal region, which we denote as region III, defined by $1.005 \gevcc < m_{KK} < 1.034 \gevcc$.\\ 

\begin{table}[h]
  \centering
  \caption{Number of observed signal events, $N_\text{sig}$, and signal efficiency, $\epsilon$, for $\BLcpKK$ decays. The regions are defined by the following invariant mass ranges -- I: $3.225 \gevcc < m_{\Lcp \antiproton} \leq 3.475 \gevcc$, II: $3.475 \gevcc < m_{\Lcp \antiproton} \leq 4.225 \gevcc$.}
  \begin{tabular}{lcc}
    \toprule
    Region & $N_\text{sig}$  &  $\epsilon$ \\
    \hline
    I&$37.7 \pm  8.0$& $(10.93\pm 0.08)\%$\\
    II &$28.2 \pm 8.4$& $(11.47 \pm 0.07)\%$\\
    \hline
    \hline
  \end{tabular}%
  \label{tab:eff}%
\end{table}%

We consider systematic uncertainties associated with the initial number of \BB pairs, the tracking efficiency, the particle identification efficiency, the limited number of MC events, the description of the background, and the description of the signal (Table~\ref{tab:syst}).

The uncertainty for the number of \BB pairs is $0.6\%$ \cite{Lees:2013rw}. We determine the systematic uncertainty for the charged-particle reconstruction to be $1.3\%$ and for the charged-particle identification (ID) to be $5.6\%$. The uncertainty for the charged-particle identification is evaluated by adding the uncertainty of the identification for each particle in quadrature. For the kaon the uncertainty is $5.6\%$, for the proton $0.7\%$, and for the pion $0.2\%$. The information on the detector-related uncertainties is described in Ref.~\cite{ref:NIM2}. The statistical uncertainty associated with the MC sample is $0.4\%$. The systematic uncertainties arising from the fit procedure are determined by changing the background description for $\mb$ from a first-order polynomial to a second-order polynomial and by changing the fit ranges in $\mes$ and $\mb$ while using a first-order polynomial for $\mb$ ($7.0\%$). Changing the signal description for $\mb$ and $\mes$ from a sum of two Gaussian functions with fixed shape parameters to a single Gaussian function whose  parameters are determined in the maximum likelihood fit leads to an uncertainty of $3.1\%$. The total systematic uncertainty is $9.6\%$, obtained by adding all contributions in quadrature. 

The $26\%$ uncertainty of the $\Lcp$ branching fraction is listed as a third uncertainty, separate from the statistical and systematic components. To be consistent with prior branching fraction measurements of baryonic \B decays, we use the current value for $\BR(\Lcp \to \proton \Km \pip)$ \cite{ref:PDG} and do not incorporate the recent measurement by Belle \cite{Zupanc:2013iki}.

Only additive systematic uncertainties, i.e., uncertainties influencing the signal and background yields differently, affect the significance of the signal. The significance of the $\BLcpKK$ signal taking into account the additive systematic uncertainties is $5.0$ standard deviations.\\

\begin{table}[b]
  \centering
  \caption{Summary of the systematic uncertainties for \BLcpKK.}
  \begin{tabular}{lc}
    \toprule
    Source & Relative uncertainty \\
    \hline
    Multiplicative uncertainties:\\
    \hspace{0.1cm} \BB counting & $0.6\%$\\
    \hspace{0.1cm} Track reconstruction & $1.3\%$ \\
    \hspace{0.1cm} Charged particle ID & $5.6\%$  \\
    \hspace{0.1cm} MC sample size & $0.4\%$ \\
    \hline
    Additive uncertainties:\\
    \hspace{0.1cm} Background description& $7.0\%$ \\
    \hspace{0.1cm} Signal description & $3.1\%$\\
    \hline
    Total & $9.6\%$ \\
    \hline
    \hline
  \end{tabular}%
  \label{tab:syst}%
\end{table}%

To determine the branching fraction, we use the following relation:
\begin{equation}
\begin{split}
&\BR(\BLcpKK) =\\
& \quad  \frac{1}{\BR(\Lcp \to \proton \Km \pip)} \cdot \frac{1}{N_{\B}}  \cdot \left( \frac{N_\text{sigI}}{\epsilon_{I}} +  \frac{N_\text{sigII}}{\epsilon_{II}} \right). 
\end{split}
\end{equation}

\noindent Here, $N_{\B} = (471\pm 3) \times 10^6$ is the initial number of \BB events \cite{Lees:2013rw}. We assume equal production of \BzBzb and \BpBm pairs. The $\Lcp$ branching fraction is $\BR(\Lcp \to \proton \Km \pip) = (5.0\pm1.3)\%$ \cite{ref:PDG}, and $N_{\text{sigI}}$, $N_{\text{sigII}}$, and $\epsilon_{I}$, $\epsilon_{II}$ are the numbers of signal events and the efficiencies in the two regions of the baryon-antibaryon invariant mass. We obtain
\begin{equation}
\begin{split}
&\BR\left(\BLcpKK \right) =\\
&\quad \left(2.5 \pm 0.4_{(\text{stat})} \pm 0.2_{(\text{syst})}\pm 0.6_{\left(\Lcp\right)} \right) \times 10^{-5}.
\end{split}
\end{equation}
\noindent Eliminating the uncertainty of the $\Lcp$ branching fraction, the result is
\begin{equation}
\begin{split}
&\BR\left(\BLcpKK \right) =\\
&\quad \left(2.5 \pm 0.4_{(\text{stat})} \pm 0.2_{(\text{syst})} \right) \times 10^{-5}\\
&\quad \times \frac{0.050}{\BR\left(\Lcp \to \proton \Km \pip \right)}.
\end{split}
\end{equation}

\noindent This result is a factor of $47$ smaller than the \BLcppipi branching fraction.

All Feynman diagram contributions for \BLcpKK lead to Feynman diagram contributions for \BLcppipi through replacement of the $\ssbar$ pair in the final state with a $\ddbar$ pair. The expectation from hadronization models for these common processes is that the \BLcppipi and \BLcpKK branching fractions should differ by a factor of 3. The expected \BLcppipi branching fraction arising from these common processes is about $7.5 \times 10^{-5}$, representing only $6.4\%$ of the observed \BLcppipi branching fraction \cite{ref:PDG}. The remaining contributions arise from other Feynman diagrams, notably diagrams with external $W$ boson emission (operator product expansion operator 1 \cite{Wilson:1969zs}), which are not allowed for \BLcpKK. Moreover, \BLcppipi decays receive a large contribution from resonant subchannels. These differences likely explain why we find the \BLcpKK and \BLcppipi branching fractions to differ more than the naive factor of 3.

\begin{figure}[t]
  \begin{center}
	\includegraphics[width=0.49\textwidth]{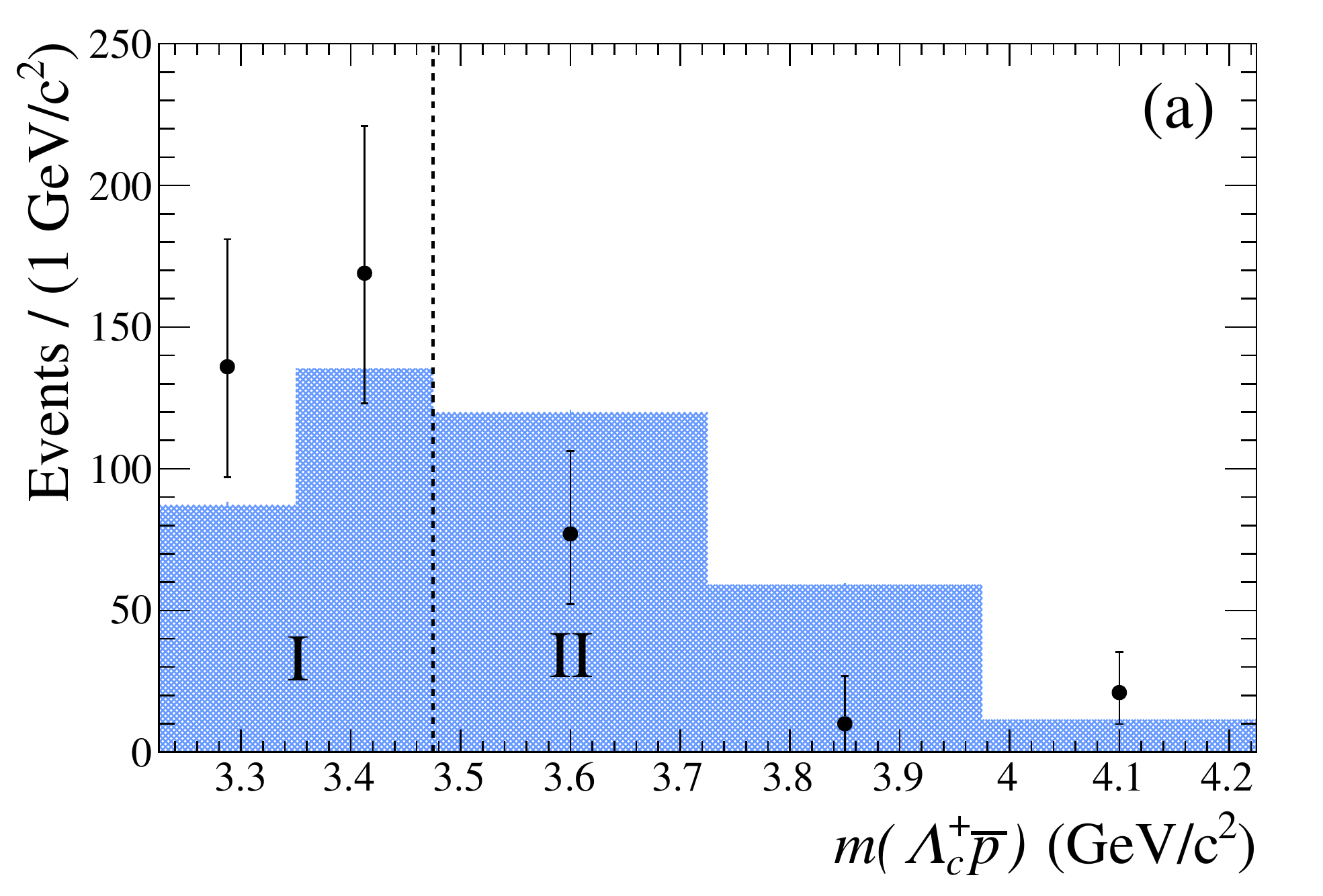}
	\hfill
	\includegraphics[width=0.49\textwidth]{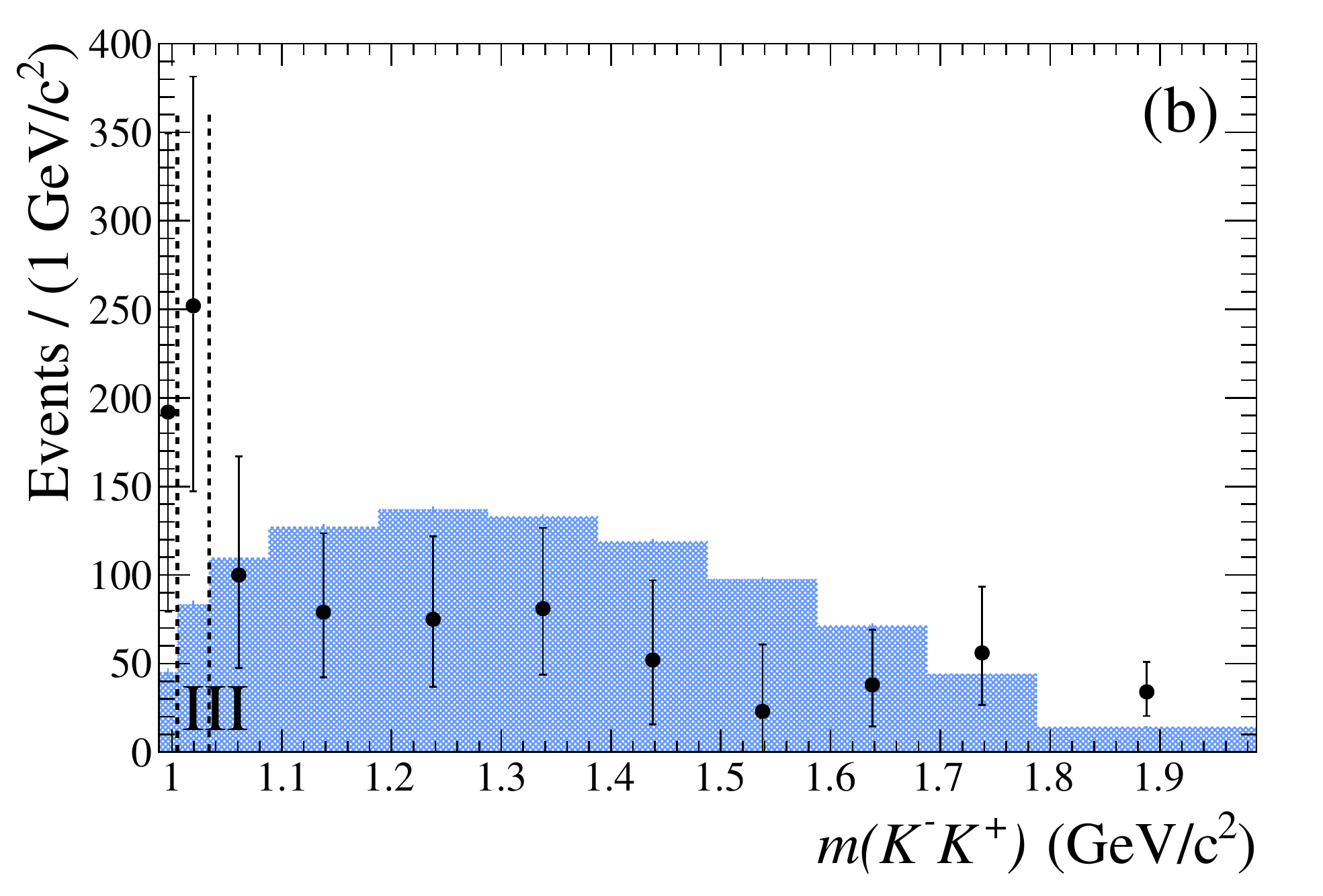} 
  \end{center}
  \caption{(a) Baryon-antibaryon invariant mass signal distribution and (b) kaon-kaon invariant mass signal distribution for data (points with statistical uncertainties) compared to distributions for simulated $\BLcpKK$ decays generated according to four-body phase space (shaded histogram), scaled to the same number of events as in data. Regions I, II, and III are indicated in the figure and described in the text.}
  \label{fig:m_Lcpbkgsub}
\end{figure}

We perform a fit in intervals of $m(\Lcp \antiproton)$ to determine the dependence of the number of signal events on the baryon-antibaryon invariant mass. The lower limit of the mass range is given by the kinematic threshold for $\Lcp \antiproton$ production, while the upper limit corresponds to the threshold $\Km\Kp$ mass with the $\Km\Kp$ system at rest in the $\Bzb$ rest frame. The results are shown in Fig.~\ref{fig:m_Lcpbkgsub}(a). The trend of the data is consistent with a small threshold enhancement, but the result is not statistically significant. The fit results for the intervals I and II in $m_{\Lcp \antiproton}$ and the detection efficiencies for these regions are shown in Table \ref{tab:eff}.

We also perform fits in intervals of $m(\Km\Kp)$. As can be seen in Fig.~\ref{fig:m_Lcpbkgsub}(b), the data deviate from the phase space expectation near threshold, in the region of the $\phi$ meson resonance. The events, in region III, include contributions from $\BLcpKK$ and $\BLcpphi$. The number of events in region III is used to determine a Bayesian upper limit at $90\%$ confidence level for the decay $\BLcpphi$ by integrating the likelihood function. This upper limit is estimated to be 17 events. The efficiency for $\BLcpphi$ decays is $(12.04 \pm 0.06)\%$. Using the result $\BR(\phi \to \Kp \Km) = (48.9 \pm 0.5)\%$ \cite{ref:PDG}, we obtain
\begin{equation} 
\BR\left(\BLcpphi \right) < 1.2 \times 10^{-5}.
\end{equation}

In summary, we observe the baryonic decay \BLcpKK with a significance of $5.0$ standard deviations including statistical and systematic uncertainties and determine the branching fraction to be $\left(2.5 \pm 0.4_{(\text{stat})} \pm 0.2_{(\text{syst})}\pm 0.6_{\BR\left(\Lcp\right)} \right) \times 10^{-5}$. The uncertainties are statistical, systematic, and due to the uncertainty in the $\Lcp \to \proton \Km \pip$ branching fraction, respectively. We obtain an upper limit of $1.2 \times 10^{-5}$ at $90\%$ confidence level for the resonant decay \BLcpphi.\\

%\input pubboard/acknow_PRL.tex
% or this one for PRDs
We are grateful for the 
extraordinary contributions of our \pep2\ colleagues in
achieving the excellent luminosity and machine conditions
that have made this work possible.
The success of this project also relies critically on the 
expertise and dedication of the computing organizations that 
support \babar.
The collaborating institutions wish to thank 
SLAC for its support and the kind hospitality extended to them. 
This work is supported by the
US Department of Energy
and National Science Foundation, the
Natural Sciences and Engineering Research Council (Canada),
the Commissariat \`a l'Energie Atomique and
Institut National de Physique Nucl\'eaire et de Physique des Particules
(France), the
Bundesministerium f\"ur Bildung und Forschung and
Deutsche Forschungsgemeinschaft
(Germany), the
Istituto Nazionale di Fisica Nucleare (Italy),
the Foundation for Fundamental Research on Matter (The Netherlands),
the Research Council of Norway, the
Ministry of Education and Science of the Russian Federation, 
Ministerio de Ciencia e Innovaci\'on (Spain), and the
Science and Technology Facilities Council (United Kingdom).
Individuals have received support from 
the Marie-Curie IEF program (European Union), the A. P. Sloan Foundation (USA), 
and the Binational Science Foundation (USA-Israel).

\end{document}